\documentclass[amsmath,amssymb,twocolumn,notitlepage,footinbib,superscriptaddress,floatfix,eprint,aps,physrev,10pt,tightenlines]{revtex4-2}

\PassOptionsToPackage{pdftex,
pdfencoding=auto,
pdfnewwindow=true,
pdfusetitle=true,
bookmarks=true,
bookmarksnumbered=true,
bookmarksopen=true,
pdfpagemode=UseThumbs,
bookmarksopenlevel=1,
pdfpagelabels=true,
breaklinks=true,
colorlinks=true,
}{hyperref}
\usepackage{hyperref}
\usepackage{amsmath, amssymb, amsthm, physics}
\usepackage[utf8]{inputenc}
\usepackage[countmax]{subfloat}
\usepackage{subfigure}
\usepackage{todonotes}
\usepackage{bbm}
\usepackage{graphicx}
\usepackage{framed}
\usepackage{color}
\usepackage{soul}
\usepackage{epstopdf}
\usepackage{dsfont}
\usepackage{relsize}
\usepackage{bm}

\footnotetext{These authors contributed equally.}

\newcommand{\pguess}{p^{\mathrm{guess}}}
\newcommand{\Q}{\mathcal{Q}}

\newcommand{\blk}{\color{black}}

\newcommand{\eqn}[1]{\begin{eqnarray} \newline #1 \end{eqnarray}}

\newcommand{\num}[1]{\begin{enumerate} #1 \end{enumerate}}
\newcommand{\hmin}{H_{\mathrm{min}}}
\newcommand{\bk}[1]{\left ( #1\right )}

\definecolor{darkgreen}{rgb}{0.0, 0.42, 0.24}

\begin{document}
\title{Experimental quantum randomness enhanced by a quantum network}

\author{Emanuele Polino$^\dagger$}
\author{Luis Villegas-Aguilar$^\dagger$}
\email{luis@villegasaguilar.com}
\affiliation{Centre for Quantum Dynamics and Centre for Quantum Computation and Communication Technology, Griffith University, Yuggera Country, Brisbane, QLD 4111, Australia}

\author{Davide Poderini$^\dagger$}
\affiliation{International Institute of Physics, Federal University of Rio Grande do Norte, 59078-970, Natal, Brazil}
\affiliation{Università degli Studi di Pavia, Dipartimento di Fisica, QUIT Group, via Bassi 6, 27100 Pavia, Italy}

\author{Nathan Walk}
\affiliation{Dahlem Center for Complex Quantum Systems, Freie Universita{\"a}t Berlin, 14195 Berlin, Germany}

\author{Farzad Ghafari}
\affiliation{Centre for Quantum Dynamics and Centre for Quantum Computation and Communication Technology, Griffith University, Yuggera Country, Brisbane, QLD 4111, Australia}

\author{Marco Túlio Quintino}
\affiliation{Sorbonne Université, CNRS, LIP6, Paris F-75005, France}

\author{Alexey Lyasota}
\author{Sven Rogge}
\affiliation{Centre for Quantum Computation and Communication Technology, School of Physics, The University of New South Wales, Sydney, NSW 2052, Australia}

\author{Rafael Chaves}
\affiliation{International Institute of Physics, Federal University of Rio Grande do Norte, 59078-970, Natal, Brazil}

\author{Geoff J. Pryde}
\affiliation{Centre for Quantum Dynamics and Centre for Quantum Computation and Communication Technology, Griffith University, Yuggera Country, Brisbane, QLD 4111, Australia}

\author{Eric G. Cavalcanti}
\affiliation{Centre for Quantum Dynamics, Griffith University, Yugambeh Country, Gold Coast, QLD 4222, Australia}

\author{Nora Tischler}
\author{Sergei Slussarenko}
\affiliation{Centre for Quantum Dynamics and Centre for Quantum Computation and Communication Technology, Griffith University, Yuggera Country, Brisbane, QLD 4111, Australia}

\begin{abstract}
The certification of randomness is essential for both fundamental science and information technologies. Unlike traditional random number generators, randomness obtained from nonlocal correlations is fundamentally guaranteed to be unpredictable. However, it is also highly susceptible to noise. Here, we show that extending the conventional bipartite Bell scenario to hybrid quantum networks---which incorporate both quantum channels and entanglement sources---enhances the robustness of certifiable randomness. Our protocol even enables randomness to be certified from Bell-local states, broadening the range of quantum states useful for this task. Through both theoretical analysis and experimental validation in a photonic network, we demonstrate enhanced performance and improved noise resilience.

\end{abstract}
\maketitle
\section{Introduction}

Quantum correlations between distant systems can defy any classical local description of reality~\cite{bell1964einstein}. These correlations, referred to as nonlocal, have not only revolutionized the way we understand the world but have also opened up new possibilities for information processing~\cite{brunner2014bell,gisin2007quantum,wehner2018quantum}.
Among these, protocols that certify randomness represent a fundamental building block for many quantum technologies, including quantum communication, simulations, and computation~\cite{barrett2005no,acin2016certified,pironio2010random,colbeck2009quantum,herrero2017quantum}.

Observing a violation of Bell inequalities implies that any model that explains these nonlocal correlations is necessarily incompatible with local causality (for careful account of all the explicit and implicit assumptions in Bell's theorem, see Refs.~\cite{wiseman2017causarum,cavalcanti2021implications}).
From a strictly operational perspective, quantum nonlocality contradicts the simultaneous acceptance of two assumptions~\cite{valentini2002signal,masanes2006general,cavalcanti2012bell}: the impossibility of faster-than-light signals, known as the principle of no-signaling, and the ability to anticipate the outcomes of a Bell test for any possible measurement, termed predictability.
Consequently, under the assumption of no-signaling, the measurement outcomes from a successful Bell test must remain unpredictable, even when conditioned on pre-existing information available to an adversarial party.
This serves as the foundation for security proofs in the context of quantum randomness certification~\cite{pironio2010random,acin2016certified}.
Importantly, this certification can be done without relying on a detailed model about the physical devices involved~\cite{colbeck2011private, pironio2013security}, in the so-called device-independent (DI) framework~\cite{scarani2012device, poderini2022ab, portmann2022security} of quantum information processing. Random numbers can be certified through tests in classical computers but with no guarantee of their privacy and security. Conversely, the DI framework can assure, under minimal assumptions, the security of the certified random numbers, granted by the validity of very general physical principles.
Randomness serves as a vital resource for applications ranging from cryptography to simulations of complex systems~\cite{acin2016certified,bera2017randomness,xu2020secure}. This drive has led to recent experimental implementations of DI and semi-DI quantum randomness certification protocols~\cite{bierhorst2018experimentally,shen2018randomness,liu2018device,zhang2020experimental,liu2021device,shalm2021device,drahi2020certified}.

A fundamental goal of DI quantum information is to develop secure protocols between two physically distant locations using entangled quantum states.
While Bell nonlocality serves as the foundational resource behind point-to-point DI protocols, it also faces significant challenges because it is susceptible to noise.
Even under optimal conditions where both parties and devices operate honestly, protocols for certifying randomness are constrained by inevitable sources of error.
Beyond certain thresholds of noise, some entangled states may admit a Local Hidden Variable (LHV) model, implying that they cannot violate any standard bipartite Bell inequality in conventional Bell tests~\cite{werner1989quantum}. Interestingly, by modifying the experimental setup, entangled states that admit an LHV model can nonetheless exhibit nonlocality, a phenomenon often referred to as activation (see Refs.~\cite{popescu1995bell,palazuelos2012superactivation, cavalcanti2011quantum,bowles2020single}).

\begin{figure}[!t]	\includegraphics[width=0.90\columnwidth]{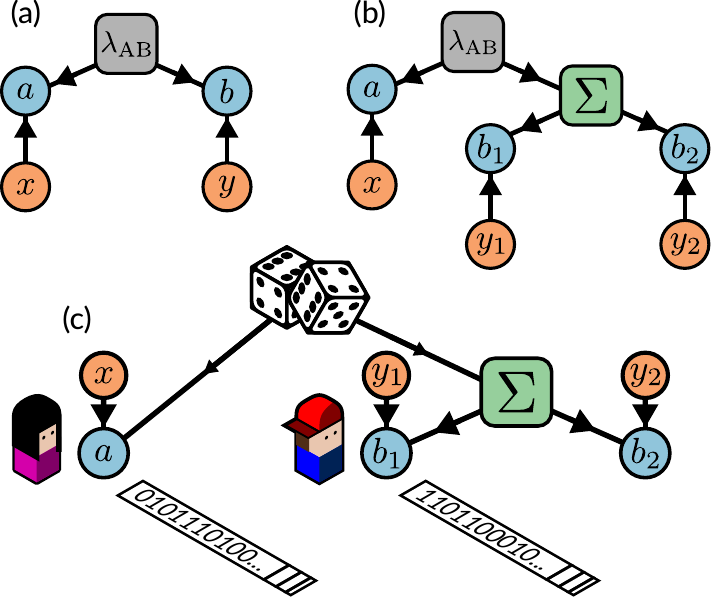}
	\caption{a) Bipartite Bell scenario. b) Broadcasting scenario. c) Randomness certification between two parties in the broadcasting scenario.}
	\label{fig:conc}
\end{figure}

Here, we theoretically and experimentally show that shared randomness derived from bipartite states can be significantly enhanced between two parties in a network, even when the original state is Bell-local.
By allowing the application of quantum operations on one of the subsystems, we transform a bipartite scenario into a multipartite one, distributing the information in the initial system amongst three parties within a network structure.
This approach can activate the nonlocality of the state under projective measurements~\cite{bowles2020single, luigi} while only using a single copy of it per experimental round, unlike other schemes that need multiple copies of the initial state distributed in networks~\cite{cavalcanti2011quantum}.
We demonstrate that such an extended network leads to both an activation of shared randomness between two parties when starting from Bell-local states and an enhancement in the randomness that can be certified for general levels of noise.
As illustrated in Fig.~\ref{fig:conc}, we focus on the scenario where two distant measurement stations wish to certify the randomness of their outputs, starting from a shared quantum resource.
However, imperfections in the physical source that produces this quantum state---or in the communication channel connecting the two stations---compromise the security of their protocol as it eventually becomes impossible to violate any standard Bell inequality between them, even if the state remains entangled~\cite{werner1989quantum}.
We experimentally construct a photonic quantum network and implement the asymmetric scenario of Fig.~\ref{fig:conc}(c), reporting a proof-of-principle demonstration of the randomness certification protocol in a four-photon experiment.
To quantify our findings, we provide upper bounds on the information that a potential adversary may obtain from the observed network statistics.

Ultimately, our work constitutes an important step toward developing randomness generation protocols that surpass the limitations of standard two-party scenarios. It extends these protocols into the domain of general network architectures that incorporate quantum sources and quantum channels, reflecting the long-term vision of quantum networks, which will involve multiple nodes in different configurations~\cite{woodhead2018randomness,grasselli2021entropy,grasselli2023boosting,wooltorton2023ExpandingBipartiteBell, li2024randomness, murta2020quantum, hahn2020anonymous, proietti2021experimental, pickston2023conference}.

\section{Randomness certification from Bell nonlocality}
\label{sec:chsh_case}

In a standard bipartite scenario, two separate parties, Alice and Bob, 
share a quantum state $\rho_\mathrm{AB}$.
Each party may perform a measurement on their physically accessible Hilbert space $H_\mathrm{A}$ or $H_\mathrm{B}$, according to classical inputs $x$ and $y$, obtaining classical outcomes $a$ and $b$, respectively.
It is assumed that their state may also be entangled with a potential adversary, Eve, whose objective is to predict  Alice's and Bob's measurement outcomes.

Formally, one considers the global state $\rho_\mathrm{ABE} \in H_\mathrm{A} \otimes H_\mathrm{B} \otimes H_\mathrm{E}$, and the reduced state shared by Alice and Bob is $\rho_\mathrm{AB} = \Tr_\mathrm{E}\left[\rho_\mathrm{ABE}\right]$.
The joint probability distribution 
is given by the Born rule $p(a,b,e|x,y) = \Tr\left[(A_{a|x}B_{b|y}E_{e}) \rho_\mathrm{ABE}\right]$, where $A_{a|x}$ denotes the Positive Operator-Valued Measure (POVM) that is associated with outcome $a$, conditioned on Alice carrying out the measurement $x$, and similarly for Bob with $B_{b|y}$.

We limit Eve to only performing individual attacks in each run of the experiment; she conducts the POVM ${E_{e}}$, obtaining outcome $e$, with the objective of correlating her outcome with those of both Alice and Bob.
Such correlations would allow Eve to predict their simultaneous outputs---a knowledge quantified by a measure known as the guessing probability
\begin{equation}
\label{eq:guessing}
    \pguess(x, y) = \max_{\{p \in \Q\}} \sum_{a, b} p(a, b, e=(a,b) | x, y) \;,
\end{equation}
where $\Q$ denotes the set of all possible probability distributions allowed by quantum theory as described above.

This quantity is, in turn, directly associated with the amount of randomness that can be certified from the observed correlations between Alice and Bob in the form of the min-entropy~\cite{konig2009operational}
\begin{equation}
    \label{eq:minentropy}
    H_{\min}^\mathrm{AB|EXY} = -\mathrm{log}_2(\pguess(x, y)) \;.
\end{equation}

From a given experimental behavior $p^{\mathrm{exp}}(a, b| x,y)$, it is possible to lower bound the values of $H_{\min}^\mathrm{AB|EXY}$ 
using linear functions of $p^{\mathrm{exp}}$, like the violation of a bipartite Bell inequality or, more
efficiently, directly from the observed probability distribution~\cite{nieto2014using}.

\section{Certified randomness in a broadcasting network}
To generate DI randomness in the standard two-party scenario in Fig.~\ref{fig:conc}(a), Alice and Bob must observe Bell nonlocal correlations generated from their shared entangled state (although nonlocality and DI randomness are, in general, different resources~\cite{ramanathan2024maximum}).
However, not all entangled states can violate a bipartite Bell inequality.
Among these, a prominent class of mixed entangled states are the two-qubit isotropic states~\cite{werner1989quantum}:
\begin{equation}
    W^{\alpha}_\mathrm{AB} = \alpha \ket{\Phi^+}\bra{\Phi^+} + (1-\alpha) \; \mathbbm{1}/4 \;,
     \label{eq:werner}
\end{equation}
where $\ket{\Phi^+}=(\ket{00}+\ket{11})/\sqrt{2}$ is a maximally entangled state and $\mathbbm{1}/4$ is a completely mixed one.

In the range $\alpha\in (1/3, 0.6875]$, these states are entangled, but do not exhibit standard Bell nonlocality under dichotomic measurements~\cite{hirsch2017better, designolle2023ImprovedLocalModelsa}.
Nevertheless, it is now known that the single-copy nonlocality of these states can be activated~\cite{bowles2020single,luigi} when the scenario is modified as in Fig.~\ref{fig:conc}(b).
The crucial step for activation in this scenario is the application of a broadcast channel $\Sigma$ on Bob's system:
\begin{equation}
 \rho_\mathrm{AB}  \xrightarrow[\text{broadcast}]{\Sigma} \rho_\mathrm{A B_1 B_2}\; .
\end{equation}
This operation entangles Bob's original particle with an ancillary qubit, creating two spatially separated subsystems $\mathrm{B_1}$ and $\mathrm{B_2}$, mapping the original bipartite scenario of Fig.~\ref{fig:conc}(a) to the tripartite one in Fig.~\ref{fig:conc}(b).

Following this channel, the entanglement initially shared between Alice and Bob is now distributed among three parties (Alice, Bob$_1$ and Bob$_2$).
Interestingly, in this asymmetric scenario, the state in Eq.~\eqref{eq:werner} can violate a suitable 
broadcasting inequality $\mathcal{I}$ tailored to this causal structure whenever $\alpha>1/\sqrt{3}$ (see Ref.~\cite{bowles2020single} and the Appendix \ref{app:broadineq}), even if it remains Bell-local for any standard Bell experiment involving projective measurements.

\begin{figure*}[ht]
  \centering
  \includegraphics[width=0.9\textwidth]{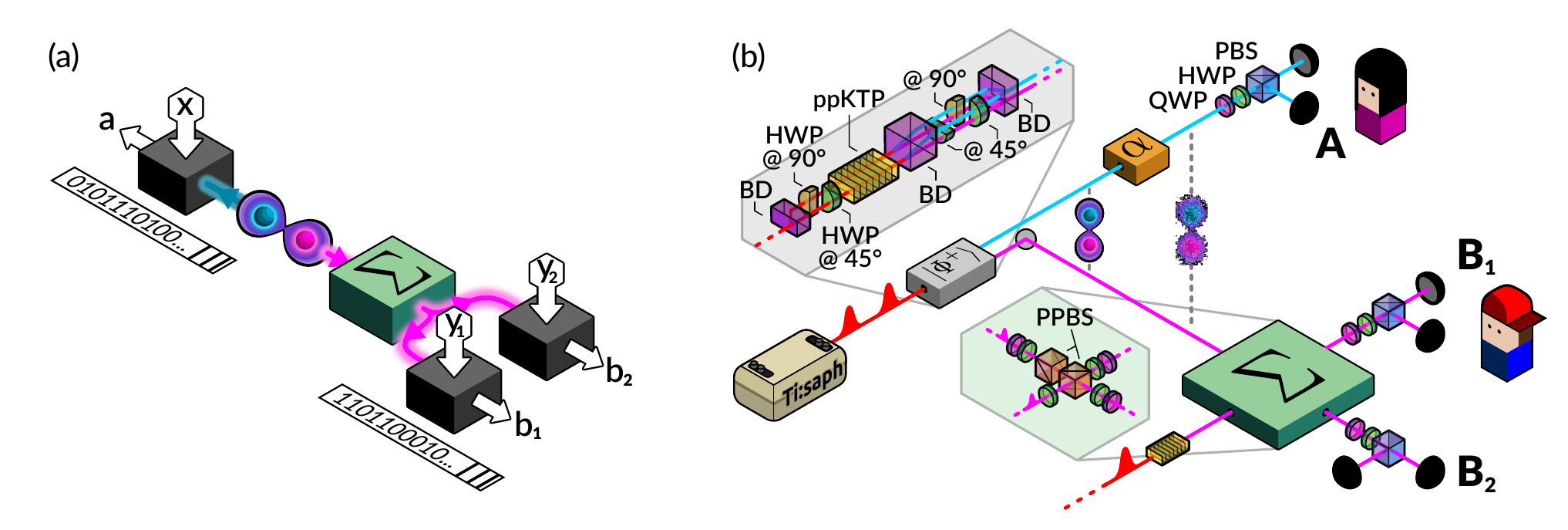}
\centering
\caption[]{\textbf{Experimental setup.}
\textbf{a)} Conceptual layout of the network-enhanced two-party randomness. A noisy bipartite state is distributed among three separate users who locally measure their photon based on inputs $x$, $y_1$, $y_2$. For two users, shared randomness is certified.
\textbf{b)}
The polarization-entangled photons originally shared between Alice and Bob are created by pumping a periodically poled potassium titanyl phosphate (ppKTP) crystal with a 775 nm pulsed laser. One of the photons is controllably depolarized, resulting in a two-qubit isotropic state with pure-state proportion $\alpha$. The other photon is probabilistically broadcast via the channel $\Sigma$ to parties ${\rm B_1}$ and ${\rm B_2}$. A second ppKTP crystal is pumped to generate the ancilla photon required for $\Sigma$ and a heralding signal. QWP, quarter-wave plate; HWP, half-wave plate; BD, beam displacer; PBS, polarizing beam splitter; PPBS, partially polarizing beam splitter.}
\label{fig:experiment}
\end{figure*}

\section{Results}
\blk
\subsection{Certified two-party randomness from the network statistics}
The detection of nonlocal correlations in the network structure of Fig.~\ref{fig:conc}(b) guides the search for quantum distributions that can certify randomness against potential adversaries---particularly when the original state may be useless for this task in a bipartite scenario. This concept is depicted in Fig.~\ref{fig:conc}(c).
To quantify the randomness that can be certified in this scenario, we follow the same spirit as in standard Bell experiments and consider an outer approximation~\cite{navascues2008convergent} of what we term the ``quantum broadcasting scenario". In practice, to establish bounds on the randomness, we impose fully quantum constraints on a larger quantum state that could be shared among Alice, Bob$_1$, Bob$_2$, and a potential adversary, Eve.

Here, we are interested in the randomness that can be certified between two parties, Alice and Bob$_1$.
Similarly to a bipartite Bell scenario, the maximum knowledge that Eve can have about their respective measurement outcomes $a$ and $b_1$, conditioned on measurement choices $x$ and $y_1$, is given by the two-party guessing probability. In the quantum broadcast scenario, this quantity is defined as:
\begin{equation}
    p^{\mathrm{guess}}_2(x,y_1) = \max_{\{p \in \mathcal{Q}\}} \sum_{a, b_1,
    b_2} p(a,b_1,b_2,e=(a,b_1) | x,y_1,y_2).
    \label{eq:2partyguess}
\end{equation}

To bound this guessing probability, one can follow either one of two standard approaches. 
The first is to derive it directly from the observed value of the broadcast inequality $\mathcal{I}$ (see Appendix \ref{app:broadineq}). 
Let $p^{\mathrm{exp}}(a,b_1,b_2| x,y_1,y_2)$ be an experimental probability distribution that produces a value $\mathcal{I}\left[p^{\mathrm{exp}}\right]$.
We can then obtain an upper bound on $p^{\mathrm{guess}}_2(x,y_1)$ compatible with $\mathcal{I}\left[p^{\mathrm{exp}}\right]$ and 
the quantum theory set $\mathcal{Q}$ by solving the following optimization problem~\cite{navascues2007bounding}:
\begin{equation}
\begin{split}
    \max~ & p^\mathrm{guess}(x,y_1)\\
    \textrm{s.t.} ~ & p(a,b_1,b_2,e|x,y_1,y_2) \in \mathcal{Q}_k,\\
    & \mathcal{I}\left[p\right] = \mathcal{I}\left[p^{\mathrm{exp}}\right]   \;.
    \end{split}
\label{sdp_viol2}
\end{equation}

This can be approximately solved using the Navascu\'es-Pironio-Ac\'in (NPA) hierarchy~\cite{navascues2007bounding}, which converts the problem of belonging to $\Q$ to a (possibly infinite) hierarchy of conditions, each expressible by a semi-definite program (SDP) optimization.
While the dimensionality of these SDPs increases with the hierarchy level $k$, each optimization constrains the probability $p$ to a set $\Q_k$ such that $\Q \subseteq \Q_k$, effectively providing an upper bound for the guessing probability.

A second approach to bound the guessing probability is to impose the complete statistics $p^{\mathrm{exp}}(a, b_1, b_2| x, y_1, y_2)$ as a constraint in the numerical optimization, as shown by Ref.~\cite{nieto2014using} for the Bell scenario. Under this constraint, we are led to a second optimization problem:
\begin{equation}
\begin{split}
    \max~ & p^\mathrm{guess}(x,y_1)\\ 
    \textrm{s.t.} ~ & p(a,b_1,b_2,e|x,y_1,y_2) \in \mathcal{Q}_k\\
    & \sum_{e} p(a,b_1,b_2,e|x,y_1,y_2) = p^{\mathrm{exp}}(a,b_1,b_2|x,y_1,y_2) \; .
    \end{split}
\label{sdp_prob2}
\end{equation}

A numerical solution for either approach, therefore, yields a valid upper bound on the maximum value of $p_2^\mathrm{guess}(x,y_1)$ from Eq.~\eqref{eq:2partyguess} that a quantum eavesdropper may have when using individual attacks. We compute both of these bounds using a level-4 NPA approach and semidefinite programming ~\cite{navascues2008convergent, WoodheadGit}.

\subsection{Experiment}

\begin{figure}[ht]
  \centering
  \includegraphics[width=0.95\columnwidth]{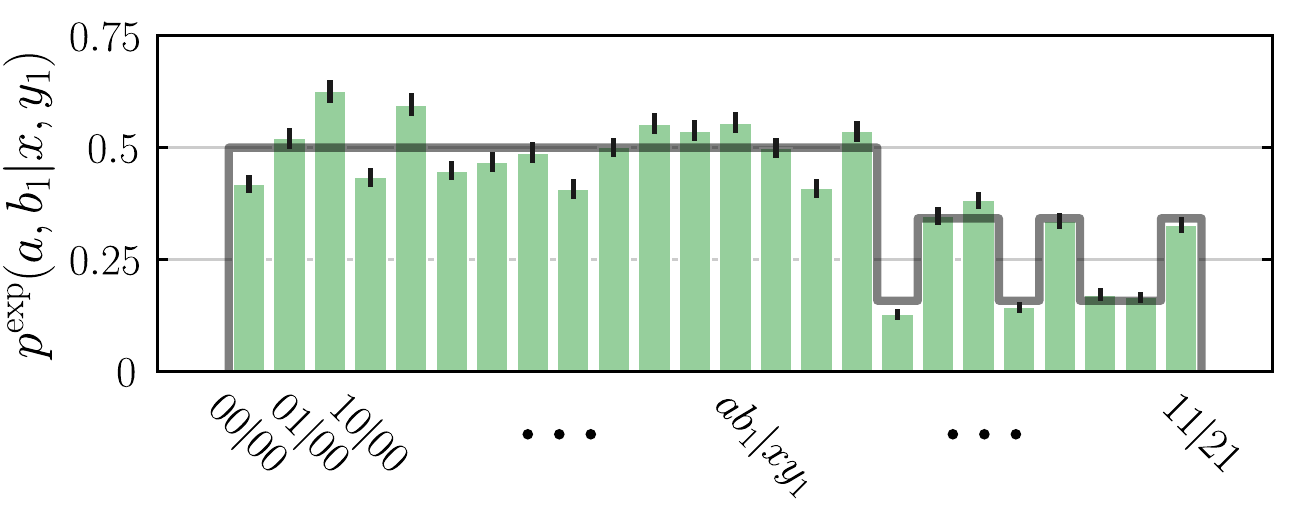}
\centering
\caption[]{\textbf{Experimental statistics.}
Experimental joint probability distribution $p^\mathrm{exp}(a,b_1|x,y_1)$ derived from measurements on the experimental state $\rho^\alpha_\mathrm{AB}$ with $\alpha = 0.637\pm 0.004$ following the application of the broadcast channel. Green bars represent the experimentally observed probabilities, while the grey line shows the theoretical distribution from an ideal network. Errors are estimated from the Poissonian statistics of the counts.}
\label{fig:pxy}
\end{figure}

In our experiment (Fig.~\ref{fig:experiment}), we implement a photonic version of the three-party network of Fig.~\ref{fig:conc}(c).
Two type-II spontaneous parametric down-conversion (SPDC) sources supply the entangled photon states of Eq.~\eqref{eq:werner} and the ancillary resources for the nondeterministic broadcasting channel $\Sigma$. We encode our qubits in the polarization state of single photons, such that $\ket{0} \equiv \ket{H}$ and $\ket{1} \equiv \ket{V}$.
The first SPDC pair source prepares the maximally entangled two-photon state $|\Phi^+\rangle$.
One of these photons is locally depolarized with a controllable noise channel $\alpha$, while the second one is used as an input to the broadcast channel $\Sigma$.
We prepare five mixed entangled states $\rho^\alpha_\mathrm{exp}$ for varying $\alpha$, each with fidelities $\mathcal{F}>0.991 \pm 0.006$ relative to the corresponding theoretical target $W^\alpha_\mathrm{AB}$ (see Ref.~\cite{luigi}).
For $\alpha \leq 0.6875$, the two-party isotropic states of Eq.~\eqref{eq:werner} are Bell-local in bipartite scenarios under projective measurements~\cite{hirsch2017better, designolle2023ImprovedLocalModelsa}.
With the algorithm of Ref.~\cite{luigi}, we show that our experimental states $\rho^\alpha_\mathrm{AB}$ with $\alpha = 0.637\pm 0.004$ are also local in the same scenarios, despite not being exact isotropic states due to experimental imperfections.
A second SPDC pair source generates a heralded ancillary photon used as a resource for the broadcast channel $\Sigma$, which includes a probabilistic photonic controlled-NOT gate~\cite{obrien2003DemonstrationAllopticalQuantum} that broadcasts half of $\rho^\alpha_\mathrm{AB}$ to two spatially separated parties, $\mathrm{B_1}$ and $\mathrm{B_2}$, with a success probability $1/6$.

The protocol begins with a sequence of experimental trials, where each party selects between two possible measurement settings ($y_1,y_2\in\{0,1\}$) in the case of Bob$_1$ and Bob$_2$, and three ($x\in\{0,1,2\}$) for Alice.
All three parties then perform a projective measurement of the polarization state of their photon. We record each of their measurement outcomes and, upon simultaneous detection of the heralding signal, count them into the three-party statistics.
In Fig.~\ref{fig:pxy}, we present the marginal bipartite probability distribution $p^\text{exp}(a,b_1|x,y_1)$ between Alice and Bob$_1$, originating from the experimental state $\rho^{0.637}_\mathrm{exp}$ after the broadcast channel, alongside ideal theoretical predictions.

\begin{figure}[ht]
  \centering
  \includegraphics[width=0.95\columnwidth]{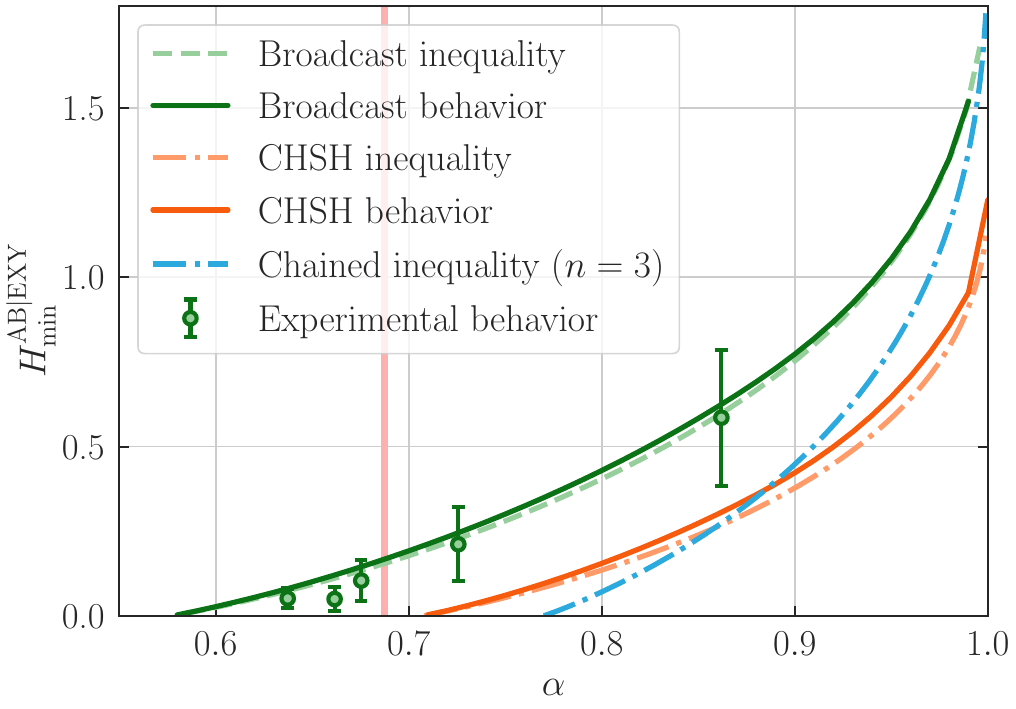}
\caption[]{\textbf{Comparison of the two-outcome min-entropy as a function of the noise parameter $\alpha$.}
Our two sets of results optimize $H_{\min}^\mathrm{AB|EXY}$ constrained on the violation of the broadcast inequality (dashed green) and full probability distribution (solid green). We compare against the two-input, two-output CHSH inequality (dash-dotted orange) and full probability distribution (solid orange), as well as the chained Bell inequality for $n=3$ (dash-dotted blue).  For $\alpha\leq 0.6875$ (vertical red line), two-party isotropic states $W_\alpha$ are Bell-local under projective measurements.
The results for the randomness certified from the experimental data (green circles) are computed from the complete experimental statistics. Error bars are calculated by considering the Poissonian errors of the events and represent $\pm 1$ standard deviation.
}
\label{fig:2parties}
\end{figure}

The results of the two-party randomness from the optimization in Eqs.~\eqref{sdp_viol2} and~\eqref{sdp_prob2} are presented in Fig.~\ref{fig:2parties} as a function of the state parameter $\alpha$.
First, we theoretically compute the achievable bounds on the certified random bits, given as the min-entropy of Eq.~\eqref{eq:minentropy}.
We constrain this value either by the estimated violation of the broadcast inequality (dashed green curve) for a state $W^\alpha_\mathrm{AB}$ or by considering the entire probability distribution (i.e., the behaviour) generated by the three parties (solid green curve). 
As expected, we observe an advantage in the produced randomness when using the complete set of outcome probabilities in the optimization program.
For the experimental data, we optimize the min-entropy using the complete set of collected statistics (green circles).

We benchmark our results with the randomness that can be certified from standard bipartite inequalies~\cite{clauser1969proposed, wooltorton2022tight, xiao2023device}.
Compared with these two-party scenarios, the immediate observation is that our results enable randomness certification in previously inaccessible noise regimes.
Notably, we achieve nonzero certifiable randomness between Alice and Bob even for states incapable of violating any standard Bell inequality (Fig.~\ref{fig:2parties}, vertical red line) under two-outcome measurements~\cite{designolle2023ImprovedLocalModelsa, luigi}.
Additionally, the two-party min-entropy achievable in our setup is larger than in the bipartite Bell scenarios considered, particularly as the noise in the system increases.
We also consider the single- and three-party randomness that can be certified with our protocol in the Appendices \ref{app:oneparty} and \ref{app:3party}. 

So far, our results (and comparisons with the standard bipartite case) hold under the assumption of individual attacks, where an eavesdropper can attack each round of the protocol individually and in an i.i.d. (independent and identically distributed) manner.
Nevertheless, we show in the Appendix that recent theoretical advances in the analysis of device-independent quantum cryptography (such as Ref.~\cite{liu2021device}) can be naturally adapted to our broadcast protocol to obtain a composable, finite-size security proof against general (coherent) attacks~\cite{acin2007device,arnon2018practical}.

\section{Discussion and outlook}

The ability to certify randomness in a DI manner stands as a fundamental application in quantum technologies. For standard bipartite scenarios, this requires the violation of suitable Bell inequalities, a task significantly impacted by noise.
In our demonstration, we show an enhancement in the randomness achievable from bipartite states in a broadcast-type network~\cite{bowles2020single, luigi} in terms of noise tolerance.
As a proof-of-principle experiment, we do not close the locality loophole, and we also rely on the fair-sampling assumption, which could be avoided with sufficiently high detection and channel efficiency.
Our analysis adopts the use of the min-entropy as a quantifier of the certifiable randomness. This approach allows casting the problem in terms of straightforward numerical optimization, albeit with the tradeoff of yielding generally looser bounds.
Nevertheless, as we discuss in the Appendix \ref{app:collective}, one can adapt recent methods~\cite{Araujo_semidefinite,brownComputingConditionalEntropies2021,brownDeviceindependentLowerBounds2024} for computing tighter bounds on the von Neumann entropy.
This would improve our results further and would be particularly relevant for exploring DI randomness expansion~\cite{shalm2021device, liu2021device} in broadcast networks.

Our findings align with recent results that indicate that multipartite protocols can surpass bipartite scenarios for the purpose of DI~\cite{grasselli2023boosting, wooltorton2023ExpandingBipartiteBell} and semi-DI~\cite{li2024randomness} randomness certification.
In our case, we demonstrate that such a multipartite advantage can be achieved using a single copy of a bipartite state.

The causal structure we consider in our experiment may naturally arise in network protocols, where two distant parties can work together to produce joint randomness over noisy channels.
When one party has access to quantum resources (e.g., local entangling gates), our protocol is able to outperform standard bipartite scenarios. We show that in the extended network, even Bell-local states can be used to certify randomness in a DI manner---extending the range of bipartite states that are useful for this task.
In the context of a future quantum internet~\cite{wehner2018quantum,xu2020secure}, networks connecting multiple stations will play a crucial role.
Different topologies have been explored as extensions of the bipartite Bell scenario, ranging from networks with a single source~\cite{bancal2013definitions, coiteux2021any, mao2022test, cao2022experimental} to those involving independent sources~\cite{fritz2012BellTheoremCorrelation, poderini2020experimental, polino2023experimental,sekatski2023partial,sarkar2024network,wang2024experimental}.
Our work broadens the study of randomness certification to multipartite networks with hybrid topologies, incorporating both sources of entanglement and quantum channels~\cite{bowles2020single, boghiu2021device, chaturvedi2022extending, lobo2023certifying, rodari2023characterizing, luigi}.

\section*{ACKNOWLEDGEMENTS}
The authors thank Lynden K. Shalm for valuable discussions.
This work was supported in part by ARC Grant No.~DP210101651 (N.T., S.S., E.G.C.~and G.J.P.) and in part by ARC Grant No.~CE170100012 (S.R.~and G.J.P.).
L.V.-A.~acknowledges support by the Australian Government Research Training Program (RTP). N.T.~is a recipient of an Australian Research Council Discovery Early Career Researcher Award (DE220101082). F.G.~was supported partly by the Griffith University Postdoctoral Fellowship (GUPF\#58938). R.C.~acknowledges the Simons Foundation (Grant Number 1023171, RC) and the Brazilian National Council for Scientific and Technological Development (Grant No. 307295/2020-6). N.W.~acknowledges funding from the BMBF (QPIC-1, Pho-Quant, QR.X). D.P.~acknowledges funding from the MUR PRIN (Project 2022SW3RPY).

\newpage


\newpage

\clearpage

\onecolumngrid

\appendix

\section{Broadcast inequality and broadcast-locality}
\label{app:broadineq}
In the scenario depicted by Fig.~\ref{fig:conc}(b), the assumptions one considers (to show nonlocal correlations from the initial source) are that of \textit{broadcast-locality}~\cite{bowles2020single}, i.e., that the source $\lambda_\mathrm{AB}$ is described by an LHV model, and that the channel $\Sigma$ distributes no-signalling resources between Bob$_{1}$ and Bob$_{2}$. 
More precisely, a behavior with probabilities $p(a, b_1, b_2|x, y_1, y_2)$, is broadcast-local if it can be decomposed as
\begin{equation}
 p(a,b_1,b_2|x,y_1,y_2) = \sum_\lambda p(\lambda) p(a|x,\lambda) p^B_\mathrm{NS}(b_1,b_2|y_1,y_2,\lambda)\:,
    \label{eq:broadcast}
\end{equation}
where $p^B_{NS}(b_1,b_2|y_1,y_2,\lambda)$ is a no-signalling distribution, that is, it respects 
\begin{equation}
\begin{split}
\sum_{b_1} p_{B}(b_1, b_2|y_1, y_2,\lambda) = \sum_{b_1} p^{B}(b_1, b_2|y_1', y_2,\lambda), 
    \label{eq:NS1}
\end{split}
\end{equation}
for all $b_2,y_1,y_1',y_2,\lambda$, and it respects
\begin{equation}
    \begin{split}
    \sum_{b_2} p_{B}(b_1, b_2|y_1, y_2,\lambda) = \sum_{b_2} p^{B}(b_1, b_2|y_1, y_2',\lambda), 
        \label{eq:NS2}
    \end{split}
\end{equation}
for all $b_1,y_1,y_2,y_2',\lambda$.

Under broadcast-locality assumptions, the correlations between Alice, Bob$_1$ and Bob$_2$ satisfy the inequality~\cite{bowles2020single}:
\begin{multline}
    \mathcal{I} = \expval{A_0B^{1}_0B^{2}_0} + \expval{A_0B^{1}_1B^{2}_1} + \expval{A_1B^{1}_1B^{2}_1} - \expval{A_1B^{1}_0B^{2}_0} \\
         + \expval{A_0B^{1}_0B^{2}_1} + \expval{A_0B^{1}_1B^{2}_0} + \expval{A_1B^{1}_0B^{2}_1} - \expval{A_1B^{1}_1B^{2}_0} \\
         - 2\expval{A_2B^{1}_0} + 2\expval{A_2B^{1}_1} \leq 4,
    \label{eq:ineqbowl}
\end{multline}
where each term denotes the tripartite or bipartite correlator (equivalent to the expected value of $\pm1$ observables) for the dichotomic measurements $A_i$, $B^{1}_j$ and $B^{2}_k$ performed by each party. Here, $B^{1}_j$ denotes the $j$-th choice of measurement of Bob$_1$ (and analogously for Bob$_2$).
A violation of this inequality can be observed under suitable projective measurements for the states described in Eq.~\eqref{eq:werner} whenever $\alpha>1/\sqrt{3}$.
Also, we note that the inequality~\eqref{eq:ineqbowl} is ``at least as good'' as the bipartite CHSH inequality~\cite{clauser1969proposed}. More precisely, if Bob$_1$ deterministically outputs $+1$, i.e., $B^{1}_0=B^{2}_1=\mathbbm{1}$, and Alice's third measurement is given by $A_3=\mathbbm{1}$, it follows that $\expval{A_iB^{1}_jB^{2}_k}=\expval{A_iB^{2}_k}$ for every $i,j,k\in\{0,1\}$, and $\expval{A_2B^{1}_0}=\expval{A_2B^{1}_1}=1$. Direct calculation then shows that, when $B^{1}_0=B^{2}_1=A_3=\mathbbm{1}$, the inequality~\eqref{eq:ineqbowl} reads as
\begin{align}
    \expval{A_0B^{2}_0} + \expval{A_0B^{2}_1} + \expval{A_1B^{2}_1} - \expval{A_1B^{2}_0}\leq 2,
\end{align}
which is the CHSH inequality up to input relabelling.

Finally, we notice that in our randomness certification procedure, we do not explicitly use the broadcast-locality assumptions, but we consider a multipartite quantum scenario with the three measurement stations of the broadcasting scenario and Eve. The concept of broadcast-locality and the inequality~\eqref{eq:ineqbowl} are used as a guide to find quantum statistics capable of certifying randomness.

\section{One-party randomness certification}
\label{app:oneparty}

\begin{figure}[ht]
  \centering
  \includegraphics[width=0.5\columnwidth]{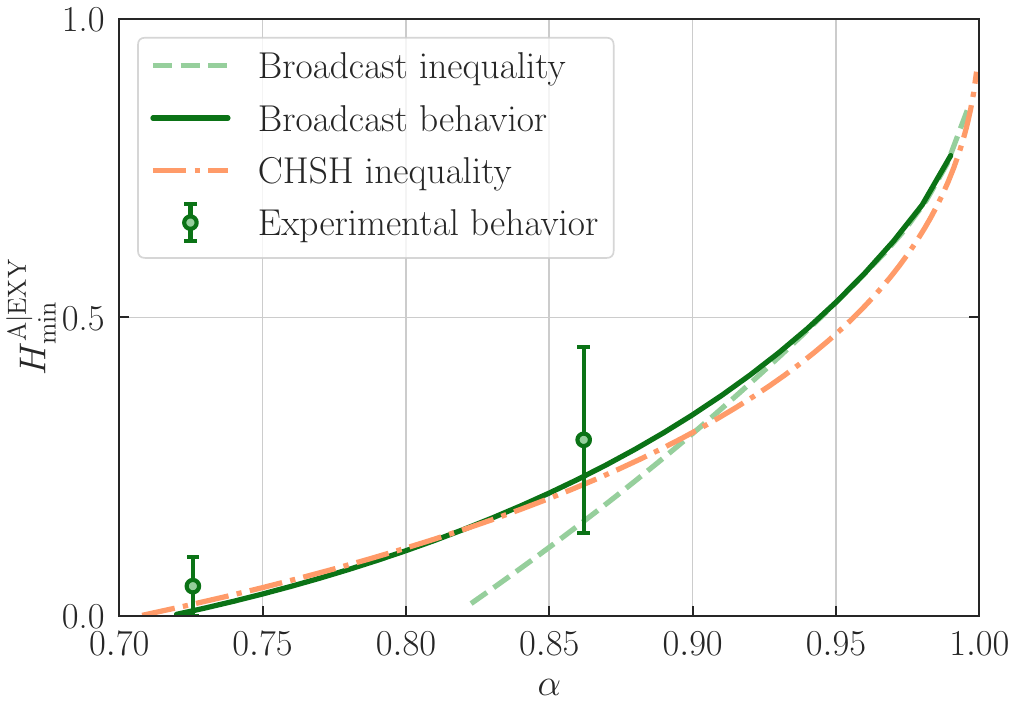}

\caption{\textbf{One-party $H_{\min}^\mathrm{A|EXY}$ as a function of the noise parameter $\alpha$.}
Results are certified using the broadcasting inequality (dashed green), constraining on the statistics for the broadcasting inequality violation (solid green), or using the CHSH inequality (dash-dotted orange). The experimental points are computed using the complete behaviour. Error bars are calculated by considering the Poissonian errors of the events and represent $\pm 1$ standard deviation.
}
\label{fig:1party}
\end{figure}

If one is interested only in the randomness of Alice's station, then the figure of merit is the one-party guessing probability:

\begin{equation}
\begin{split}
& p^{\mathrm{guess}}_1(x) = \\
& = \max_{\{p \in \mathcal{Q}\}} \sum_{a,b_1,b_2} p(a, b_1,b_2,e=a | x) \; .   
\label{eq:1partyguess}
\end{split}
\end{equation}
where there is no dependence on the input variables $y_1$ and $y_2$ because of the no-signalling conditions.
The results of the calculation and experimental results are shown in Fig.~\ref{fig:1party} and demonstrate how, for values of $\alpha > 0.83$, the broadcasting behaviors are able to improve the certified randomness with respect to the CHSH inequality. Contrary to the two-party case, the certifiable one-party randomness for high values of noise, $\alpha  \lesssim 0.83$, is greater in the standard CHSH than the broadcasting behavior.

\section{Three-party randomness certification}
\label{app:3party}

\begin{figure}[ht]
\centering
  \includegraphics[width=0.5\columnwidth]{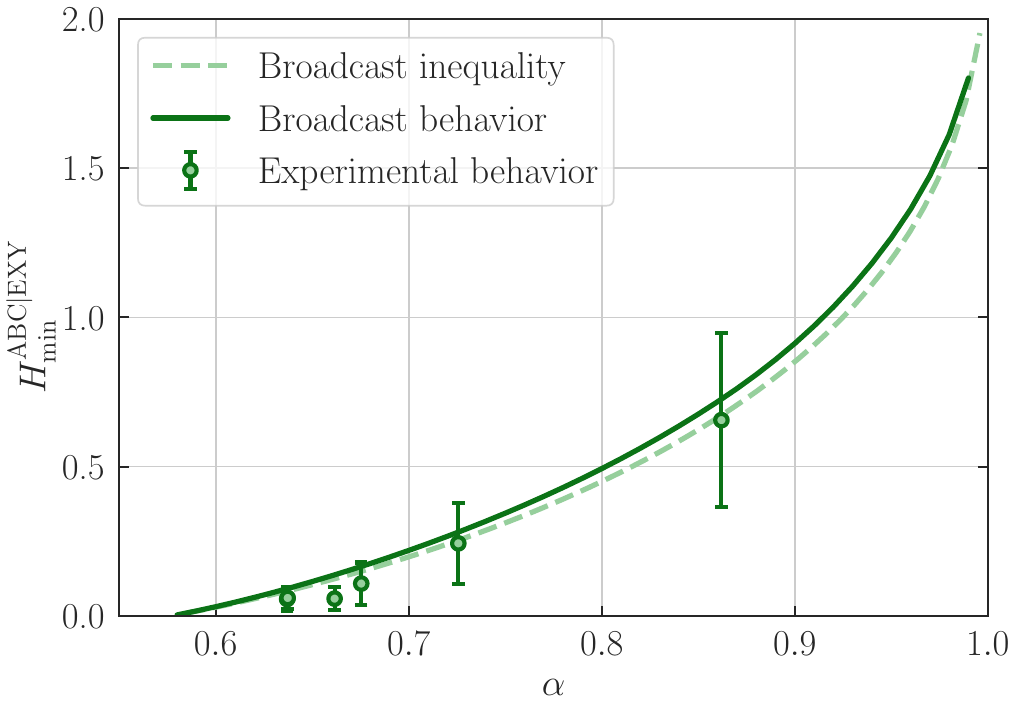}
\caption[]{\textbf{Three-party $H_{\min}^\mathrm{ABC|EXY}$ as a function of the noise parameter $\alpha$.}
Similar to Figs.~\ref{fig:2parties} and \ref{fig:1party} but for three parties, optimized over the broadcast inequality violation (dashed green) and the full measurement statistics (solid green).
Since these results depend on measurement outcomes from three distant parties, a direct comparison with the CHSH inequality is not possible.
}
\label{fig:3party}
\end{figure}

Since the broadcast network involves three independent parties, it is natural to consider the global (i.e., tripartite) randomness that can be certified from the joint statistics.
In this case, the three-party guessing probability is defined as:
\begin{equation}
\begin{split}
& p^{\mathrm{guess}}_3 (x,y_1,y_2) = \\
& = \max_{\{p \in \mathcal{Q}\}} \sum_{a,b_1,b_2} p(a, b_1,b_2,e=(a,b_1,b_2) | x,y_1,y_2) \; .   
\label{eq:3partyguess}
\end{split}
\end{equation}
The results are shown in Fig.~\ref{fig:3party} and demonstrate how three-party randomness can also be certified with broadcasting statistics. 

\section{Randomess expansion against general attacks}
\label{app:collective}
In the main text, the certifiable min-entropy generated by the network broadcast protocol is calculated and compared to the standard bipartite case under the assumption of individual attacks.
In this section, we show that the entropy accumulation theorem (EAT)~\cite{arnon2018practical,dupuisEAT2020, DupuisEATimproved2019} can be used to extend the i.i.d. security of our protocol to the general case.
More specifically, these techniques were recently applied (and slightly improved upon) to evaluate the performance of a DI randomness expansion protocol based upon the standard CHSH game~\cite{liu2021device}. 

For all of their theoretical rigour, actually evaluating the exact rates that can be achieved via the EAT is notoriously involved and typically requires extensive, sophisticated numerical optimization to obtain tight (not artificially pessimistic) rates. This would be particularly important in our case so as not to unduly penalize the broadcast network protocol but, instead, accurately identify the parameters for which a performance advantage can be demonstrated under this most general, rigorous security analysis. We expect this analysis and comparison to be highly nontrivial and, as such, defer a detailed evaluation of the performance against general attacks to future work. In what follows, we will restrict ourselves to showing how the techniques of Ref.~\cite{liu2021device} can straightforwardly be adapted to our scheme.

\subsection{Device independent randomness expansion}
The standard methods for rigorously assessing the performance and security of quantum cryptography are derived from the abstract cryptography framework~\cite{portmann2022security}. The application to the specific task of DI randomness expansion is well established (see, e.g. Supplementary Material I in Ref.~\cite{liu2021device}), but we will briefly recapitulate the essentials here. The idea is that a DI protocol, which requires some input random string of total length $\mathrm{rand_{in}}$, is executed and generates a measurement output ${\bf O}$ of length $m$ with substantial min-entropy with respect to any eavesdropper. These measurements should then be fed into a quantum-secure randomness extractor where they are processed along with a uniformly random seed $\mathbf{S}$ to produce an output ${\bf R}$ of length $\ell < m$ which should be `almost perfectly' random, which we write as $\mathbf{R} = \mathrm{Ext}(\mathbf{O},\mathbf{S})$. If this randomness extractor is also a so-called \emph{strong} extractor, then the random seed may be reused indefinitely and need not be counted in the randomness consumption of the protocol. We can say that the overall protocol has achieved net randomness expansion if $\ell - \mathrm{rand_{in}}>0$.

The notion of almost perfect randomness can be made precise as follows: we call a string ${\bf R}$ $\varepsilon$-random with respect to an observer holding a (quantum) system $E$ if
\eqn{D(\rho_{\mathbf{R}E},\tau_\mathbf{R}\otimes\rho_E)\leq \varepsilon ,}
where $D(\rho,\sigma):= 1/2 ||\rho - \sigma||$ is the trace distance and $\tau_\mathbf{R}$ is a uniformly random state on $\mathbf{R}$. The state $\tau_\mathbf{R}\otimes\rho_E$ can be interpreted as the output state of an ideal random source, i.e., a uniformly distributed string in a tensor product with the eavesdropper. The trace distance between two states has a direct relation to the optimal probability of distinguishing those states. For example, given a uniform prior over whether a real or perfect resource is prepared, then $p_{\mathrm{dist}} = \frac{1}{2} +\frac{D(\rho,\sigma)}{2} $. This allows us to interpret $\varepsilon$ as the probability that any protocol making use of $\mathbf{R}$ `fails' in the sense of producing an output that is different from that which would have been obtained using a perfect random string. Essentially, if any application using $\mathbf{R}$ produced a `bad' outcome with a probability greater than $\varepsilon$, this would be a distinguishing protocol operating with a probability greater than that which is possible given the trace distance bound, hence $\varepsilon$ is a worst-case failure probability~\cite{portmann2022security}. Returning to randomness extractors, we can now define a quantum-proof ($\ell,\varepsilon)$-strong randomness extractor as a function $\mathrm{Ext}(\mathbf{O},\mathbf{S})=\mathbf{R}$ such that
\eqn{D(\rho_{\mathbf{RS}E}, \tau_{\mathbf{R}}\otimes\tau_{\mathbf{S}}\otimes\rho_E) \leq \varepsilon.}
Note that this definition captures the notion of a strong extractor---that the seed, $\mathbf{S}$ remains close to the uniformly random state $\tau_\mathbf{S}$.

Intuitively, there should be a trade-off between the length, $\ell$, of the extracted random string and its quality as quantified by $\varepsilon_S$, and it turns out this is precisely quantified via the conditional min-entropy of the measurement outputs with respect to the eavesdropper. To be fully general, we must consider all $n$ rounds simultaneously. Let 
\eqn{\rho_{\mathbf{O}E} = \sum_\mathbf{o} p(\mathbf{o}) \ket{\mathbf{o}}\bra{\mathbf{o}} \otimes \rho_E^\mathbf{o}, \hspace{2mm} |\mathbf{o}| = n} 
be the joint state of the entire $n$-round protocol with $\rho_E^\mathbf{o}$ the state of Eve conditioned on a particular string of $n$ outputs given by {\bf o}. Then define the conditional min-entropy for the entire output as
\eqn{\hmin(\mathbf{O}|E)_{\rho_{\mathbf{O}E}}:=  -\log \max_{\mathcal{E}_{\mathbf{o}}} \left \{ \sum_{\mathbf{o}} p(\mathbf{o}) \mathrm{tr} \bk{\mathcal{E}_\mathbf{o} \rho_E^\mathbf{o}} \right \},}
where the maximization is taken of Eve's measurement operators $\mathcal{E}_{\mathbf{o}}$. From this we can recover the relation that $\hmin(\mathbf{O}|E) = -\log\bk{p^{\mathrm{guess}}}$ where $p^{\mathrm{guess}}$ is now the probability for Eve to guess the entire output string (we suppress the subscript on $\hmin(\cdot|\cdot)$ when clear from context). In the main text, we calculated this quantity for a single round, i.e. $\hmin(O^i|E)$. Under the assumption of individual attacks where Eve interacts with each round in the protocol separately, there is a tensor product structure between the rounds such that $\hmin(\mathbf{O}|E) = n \hmin(O^i|E)$. In the general case, we can use the EAT theorem to bound the min-entropy without any such assumptions, as explained in the next section.

To maximize the performance it is typically convenient to use a \emph{smoothed} version of this quantity defined as
\eqn{\hmin^\varepsilon(\mathbf{O}|E):= \max_{\sigma_{\mathbf{O}E} \in \mathcal{B}^{\varepsilon}(\rho_{\mathbf{O}E})}\hmin(\mathbf{O}|E)_{\sigma_{\mathbf{O}E}},}
where now the maximization is over all states in that $\varepsilon$ close to $\rho_{\mathbf{O}E}$ in the purified distance. Finally, it can then be proven that for an extractor utilizing two-universal hash functions (e.g. Toeplitz hashing) with a uniformly random seed and an input with smooth conditional min-entropy $\hmin^{\varepsilon_h}(\mathbf{O}|E)$ that
\begin{equation}\label{hash}
    D(\rho_{\mathbf{RS}E}, \tau_{\mathbf{R}}\otimes\tau_{\mathbf{S}}\otimes\rho_E) \leq 2^{\frac{1}{2}\bk{\ell - \hmin^{\varepsilon_h}(\mathbf{O}|E)}} + 2 \varepsilon_h : = \varepsilon_R.
\end{equation}\
In this way, knowing the smooth min-entropy makes it possible to trade off a linear reduction in the length of the extracted random string for an exponential reduction in the $\varepsilon$-randomness of that string. Now, for any desired value of $\varepsilon_R$, we simply reduce $\ell$ until the r.h.s. of Eq.~(\ref{hash}) is sufficiently small. If there is no $\ell >0$ such that this is true, then we say that no random numbers can be extracted for that $\varepsilon_R$ value.

The main task of a security analysis, therefore, boils down to bounding the multi-round min-entropy from empirically observed quantities. Quantum cryptographic protocols generally involve some kind of `spot-checking' where certain measurements (e.g.~the score in a nonlocal game) are made, and the protocol is aborted if a particular quality threshold is not achieved. If we define $\Omega$ as the event that the protocol does not abort, $p_\Omega$ probability of not aborting and $\rho_{\mathbf{R}E}$ as the output state conditioned on not aborting, then a DI randomness expansion protocol is called $\varepsilon_S$-secure (or  $\varepsilon_S$-random, or has a soundness error of $\varepsilon_S$) if
\eqn{p_\Omega D(\rho_{\mathbf{R}E},\tau_\mathbf{R}\otimes\rho_E)\leq \varepsilon_S .\label{sec}}

This represents the joint probability of all the checks being passed whilst the output string has a certain failure probability in the sense defined above. The point is that, although we cannot control (or even know) the success probability $p_\Omega$ since this can be set by the eavesdropper, we can be sure that \emph{either} the protocol aborts with high probability \emph{or} the output is very close to random with respect to the eavesdropper. Crucially, this definition is also \emph{composable}, in the sense that if multiple outputs are combined in a single application, then the $\varepsilon$ for the total protocol can be readily bounded from those of the component resources. This is particularly important for cryptographic applications where multiple resources are frequently combined. However, it should be noted that, for DI protocols, it is not currently known how to prove any protocol secure under this definition if we consider the composition of multiple protocols using the same untrusted devices. In that sense, it is only possible to prove security under the composition of output strings. Full composability for both strings and devices would require a restriction to protocols in which each device is only ever used once~\cite{portmann2022security,barrett2013memory}. 

Now, if the result of a spot-checking protocol is such that, conditioned on the quality thresholds being passed, either the protocol will abort or the smooth conditional min-entropy of the eavesdropper will be lower bounded, then this could be combined with the hashing bound in Eq.~(\ref{hash}) to certify $\varepsilon_S$-randomness protocol with  $\varepsilon_S =  \max \left \{\varepsilon_\mathrm{ENT}, 2^{\frac{1}{2}\bk{\ell - \hmin^{\varepsilon_h}(\mathbf{O}|E)}} + 2 \varepsilon_h \right \}$. In the next section, we will explain how just such a result can be obtained by extending the EAT theorem and the work of Ref.~\cite{liu2021device} to the broadcast protocol.

\subsection{Protocol, assumptions and the entropy accumulation theorem}
Following Ref.~\cite{liu2021device} we will make the following assumptions about our setup
\begin{enumerate}
\item The users possess a trusted classical computer.
\item The users have some initial trusted random strings.
\item Quantum theory is correct and complete. 
\item The users have a secure laboratory and can prevent any devices from receiving or sending communication at their discretion.
\end{enumerate}

For this last point to hold true, the devices could be either space-like separated or able to be shielded from one another. All that is different in the broadcast case is that there is a third measurement station (Bob$_2$) that must be separated/shielded in the same way Alice and Bob$_1$ are in the standard bipartite protocol. 

The protocol for broadcast-based randomness expansion then proceeds as follows:
\\

        {\bf Broadcast-network-based DIRE protocol}

        \num{\item For every round $i \in \{1,\dots, n\}$, do steps 2-4. 
        \item Choose $T^i \in \{0,1\} $ such that $p(T^i=1) = \gamma$.
        \item If $T^i = 0$, set the inputs for Alice and both Bob's ($X^i,Y_1^i,Y_2^i$ respectively) equal to 0, carry out the nonlocal game, record the output as Alice and Bob$_1$'s measurement results $O^i = A^iB_1^i$ and set $U^i = \perp$. 
        \item If $T^i=1$, draw the inputs $X^i,Y_1^i,Y_2^i$ uniformly at random from $\{0,1\}$, play the nonlocal game recording $A^i$, $B_1^i$ and $B_2^i$. If the game is won, set $U^i=1$; otherwise, set $U^i=0$. 
        \item If $|U_i: U_i = 0| > n\gamma (\omega_{\mathrm{exp}} - \delta)$, then abort the protocol.
        \item Apply a strong quantum-proof randomness extractor to get output randomness {\bf R} = Ext({\bf AB},  {\bf S}). (Because we use a strong extractor,  {\bf R} can be concatenated with  {\bf S} to give  {\bf M} = ({\bf R},  {\bf S}). }
One could also phrase the abort condition in terms of a threshold of the nonlocal score $\mathcal{I}$ of Eq.~\eqref{eq:ineqbowl}.
In one round of the protocol, the final state (after all steps have been executed) could then be written as a classical quantum state $\rho_{A^1B_1^1B_2^1X^1Y_1^1Y_2^1U^1R^1_AR^1_{B_1}R^1_{B_2}E}$ where we have included all of the classical registers as well as the post-measurement quantum states of all legitimate parties $\rho_{R_{A}}$ etc.~and the eavesdropper. Defining the input and output registers as $I^1 = X^1Y_1^1Y_2^1$ and $O^1 = A^1B_1^1B_2^1$, respectively, and the post-measurement states of the legitimate parties as $R^1 = R^1_AR^1_{B_1}R^1_{B_2}$, we could also write this as $\rho_{O^1I^1R^1E}$. Analysis of a general $n$-round protocol via the entropy accumulation theorem (EAT) proceeds by considering each subsequent round as being generated from the previous one by a set of channels $\{M_i\}$ where $M_i: R^{i-1}\rightarrow O^iI^iU^iR^i$ maps the quantum state from the previous round into the state and classical registers for the current round. This means we can write the final state of an $n$-round protocol as
\eqn{\rho^{(n)}_{\mathbf{OIU}RE} = (M_n\circ M_{n-1}\circ \dots \circ M_{1})\rho_{R^0E},}
where we use boldface to denote the $n$-length registers for the inputs, output and scores for all rounds. If these channels obey certain properties, they are called EAT channels, and the conditional min-entropy of the total protocol can be lower bounded by a function of the von Neumann entropy that is generated in a single round (which is something we can usually calculate straightforwardly). In particular, the channels must satisfy the properties
\num{\item The registers {\bf OIU} must be classical and finite-dimensional.
\item It must hold that $I(O^{i-1}:I^i|I^{i-1}E) = 0$ $\forall$ $\rho_{R^0E}$.
}
For our protocol, the first condition is true by definition. The second is also automatically satisfied since we can choose the inputs for each round uniformly at random, so they are also independent of the outcomes of the previous round. Therefore this protocol is indeed described by EAT channels.

For EAT channels, the min-entropy for all $n$ rounds can be bounded by a function that primarily depends upon $h(\omega_{\mathrm{exp}})$, a worst-case lower bound to the von Neumann entropy generated in a single round of a protocol with a nonlocal game-winning probability $\omega_{\mathrm{exp}}$. It is possible to derive a lower bound of the form
\eqn{\hmin^{\varepsilon_h}(\mathbf{O}|\mathbf{I}E) \geq n h(\omega_{\mathrm{exp}}) - \sqrt{n} \nu \label{EAT},}
where $\nu$ is a complicated formula for various finite-size corrections. More specifically, we can now directly apply \cite[Theorem 3]{liu2021device} to get an exact expression. Now, assuming that there is an $\ell >0$ such that the Eqs.~(\ref{hash}-\ref{sec}) are satisfied, then the net randomness of the protocol will be given by

\eqn{r_{\mathrm{net}} = \ell - \mathrm{rand_{in}} = \ell- n(H_{\mathrm{bin}}(\gamma)  + 3\gamma ) +2,}
where $H_{\mathrm{bin}}$ is the binary entropy, the $H_{\mathrm{bin}}(\gamma)$ term quantifies the randomness needed to randomly choose the rounds for spot-checking and the $3\gamma$ term accounts for the fact that, in the spot-checking rounds, all 3 players need to randomly choose their measurement settings to play the nonlocal game.

We conclude with some comments regarding what further steps would be required to evaluate these expressions to obtain tight rates via the EAT theorem and exactly establish the parameter regimes for a randomness expansion advantage for the broadcast network against general adversaries. The main quantity required to evaluate Eq.~(\ref{EAT}) is a convex lower bound on the von Neumann entropy generated in a single round of the protocol. The single-round min-entropy evaluated in the main text would already serve as such a bound. However, recent results suggest that there are more optimal numerical techniques to obtain better rates~\cite{Araujo_semidefinite,brownComputingConditionalEntropies2021,brownDeviceindependentLowerBounds2024}. Once a tight convex bound is obtained, this is then turned into an affine lower bound otherwise known as a min-tradeoff function. This can then be directly plugged into \cite[Theorem 3]{liu2021device} to finally obtain the min-entropy, although there are other free parameters that must also be numerically optimized. 

\end{document}